\documentclass[12pt]{iopart}

\usepackage{graphicx}
\usepackage{iopams}

\begin{document}

\title[Ad- and desorption of Rb atoms on a gold nanofilm measured by SPP]{ Ad- and desorption of Rb atoms on a gold nanofilm measured by surface plasmon polaritons}
\author{C Stehle, H Bender, F Jessen, C Zimmermann and S Slama}
\address{Physikalisches Institut, Eberhard Karls Universit\"at T\"ubingen,
Auf der Morgenstelle 14, D-72076 T\"ubingen, Germany}
\ead{slama@pit.physik.uni-tuebingen.de}

\begin{abstract}
Hybrid quantum systems made of cold atoms near nanostructured surfaces are
expected to open up new opportunities for the construction of quantum
sensors and for quantum information. For the design of such tailored quantum
systems the interaction of alkali atoms with dielectric and metallic
surfaces is crucial and required to be understood in detail. Here, we
present real-time measurements of the adsorption and desorption of Rubidium
atoms on gold nanofilms. Surface plasmon polaritons (SPP) are excited at the
gold surface and detected in a phase sensitive way. From the temporal change
of the SPP phase the Rubidium coverage of the gold film is deduced with a
sensitivity of better than $0.3\%$ of a monolayer. By comparing the
experimental data with a Langmuir type adsorption model we obtain the
thermal desorption rate and the sticking probability. In addition, also
laser-induced desorption is observed and quantified.
\end{abstract}
\maketitle

\section{Introduction}

Surface plasmon polaritons have been used for a long time as sensitive
detectors e.g. for biomolecules. This is mainly done in the so-called
Kretschmann configuration by internal reflection of a laser beam from a
surface that is coated with a thin metal film \cite%
{kretschmann_radiative_1968}. Particles that are close to or stick to the
metal film can be detected via their optical properties which shift the
surface plasmon angle $\alpha _{\mathrm{pl}}$ under which a SPP can be
excited. In principle, such shifts can be detected very sensitively by
measuring the phase of the reflected light beam which, at the plasmon angle,
changes dramatically. However, the fact that the reflectivity is zero at the
plasmon angle, i.e the power of the laser beam is completely transferred
into a plasmonic excitation, broadens the detectable phase shift and limits
the sensitivity \cite{lipson06}. Recently, a new excitation scheme has been
introduced which avoids this problem by an electronic feedback on the phase
of the incident laser field \cite{knig_real-time_2008}. In the work
presented here, this scheme is used for the first time to actually measure
the optical properties of a sample. The work is mainly motivated by recent
progress in the field of surface quantum optics, atom chips and hybrid
quantum systems. Such tailored quantum systems made of cold atoms near
nanostructured surfaces are expected to open up new opportunities for the
construction of quantum sensors and for quantum information. In this
context, we recently studied clouds of ultracold atoms coupled to
evanescent waves near solid surfaces \cite{bender_towards_2009}. The
enhancement of such optial near-fields by the excitation of surface plasmons
is already known for some time \cite{esslinger_surface-plasmon_1993}, but
only recently, proposals have been put forward to generate plasmonic
nanopotentials for the manipulation and trapping of cold atoms \cite%
{chang_trapping_2009, murphy_electro-optical_2009}. For the design of such
traps, surface potentials such as Casimir-Polder potentials \cite%
{casimir_influence_1948} have to be taken into account and some effort has
been invested by the cold atoms community for studying dissipative forces %
\cite{sandoghdar_direct_1992, sukenik_measurement_1993,
grisenti_determination_1999, obrecht_measurement_2007, bender_direct_2010}.
However, additional potentials from particles adsorbed at the surface are
only little considered. As has been shown experimentally \cite%
{mcguirk_alkali-metal_2004}, such potentials can dramatically influence atom
traps near surfaces. Thus it is crucial to understand and control atomic
adsoption, e.g. by laser-induced atomic desorption (LIAD) \cite%
{meucci_light-induced_1994}. Furthermore, fascinating ideas have been
proposed that rely on the controlled adsorption of atoms on semiconductor
surfaces, which might prove as an important technological step towards the
manipulation of electronic nano devices \cite{judd_zone-plate_2009}.\newline

Here, we report on adsorption of thermal Rubidium atoms on a gold nanofilm.
The Rubidium coverage is monitored by the phase shift in the excitation
spectrum of surface plasmon polaritons. By rapidly changing the
Rubidium vapor pressure, the balance between adsorption and thermally
induced desorption is perturbed and reequilibration is monitored in real
time. The observations are compared to a simple Langmuir type model and the
parameters of the model are extracted from the data.

\section{Experimental setup}

\begin{figure}[th]
\centerline{\scalebox{0.8}{\includegraphics{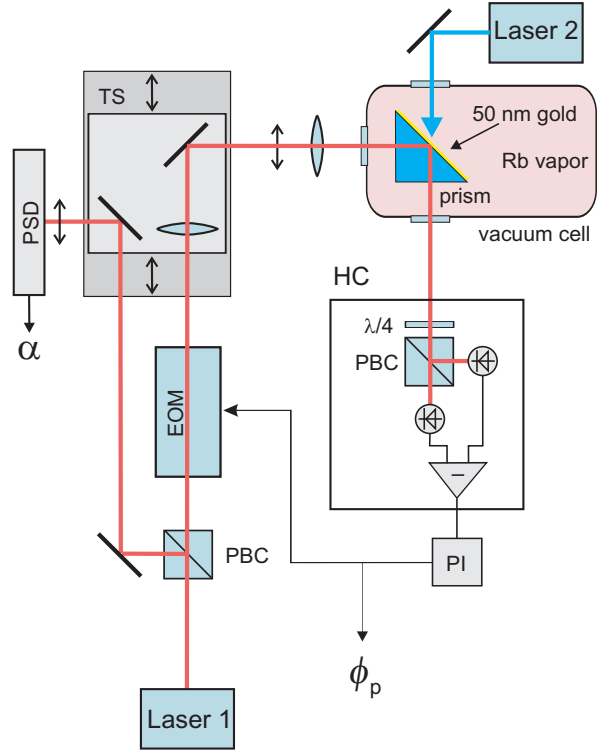}}}
\caption{ Experimental setup. Laser 1 excites plasmons at the surface of a
gold coated prism inside a vacuum chamber with a variable Rubidium (Rb)
vapor pressure. The angle $\protect\alpha $ at which the laser beam hits the
surface is varied with a translation stage (TS). It shifts the beam relative
to the lens which guides the beam to the prism surface. The position of the
translation stage is monitored with an auxiliary laser beam and a
position-sensitive diode (PSD). The beam reflected from the surface is
analyzed by a H\"{a}nsch-Couillaud (HC) setup. It consists of a $\protect%
\lambda /4$ retardation plate, a polarizing beam splitter cube (PBC) and a
pair of differential photodiodes. The recorded signal is controlled by a PI
feedback loop which acts on the phase of the incident light field via an
electro-optic modulator (EOM). The feedback-voltage is directly proportional
to the phase of the reflected light field $\protect\phi _{p}$. For
laser-assisted desorption of atoms the gold surface can be exposed to a
second laser beam (laser 2).}
\label{fig:opt_el_aufbau}
\end{figure}
The ad- and desorption of Rubidium atoms is observed by phase sensitive
detection of the surface plasmon resonance. The experimental setup is shown
in figure~\ref{fig:opt_el_aufbau}. The beam from a grating stabilized diode
laser (laser 1, wavelength near the D2-line of ${}^{87}\mathrm{Rb}$ at $%
\lambda =780$~nm, laser power $P=10$~mW, laser bandwith $2$~MHz) is internally
reflected from the gold-coated surface of a dielectric prism. The incident
angle $\alpha $ at which the laser hits the prism surface can be adjusted 
with a mirror mounted onto a mechanically driven translation stage
(TS). It shifts the beam relative to a lens which guides the beam to the
prism surface. The position of the translation stage is monitored with an
auxiliary laser beam and a position-sensitive diode (PSD). The guiding lens
is the second lens in a $2f$ telescope with the first lens being attached to
the translation stage. By that the incident beam is collimated at the prism
surface. At a characteristic plasmon angle $\alpha =\alpha _{\mathrm{pl}}$
surface plasmons are excited by the p-polarized fraction of the incoming
light. The incident light is absorbed by the plasmon excitation and the
reflectivity for p-polarized light $r_{p}$ drops to a minimum (figure~\ref%
{fig:refl_phase}). If Rubidium atoms are adsorbed, the value of the plasmon
angle shifts and the phase of the p-polarized fraction of the reflected
light field $\phi _{p}$ changes. This phase
change is monitored with a standard H\"{a}nsch Couillaud (HC) setup \cite%
{hansch_laser_1980}. A PI-servo electronics feeds the signal to an
electro-optic modulator (EOM) that controls the phase of the incident laser
light \cite{knig_real-time_2008}. This locks the H\"{a}nsch Couillaud signal
to zero such that the control-voltage at the EOM is directly proportional to 
$\phi _{p}$.\newline

\begin{figure}[tbph]
\centerline{\scalebox{0.7}{\includegraphics{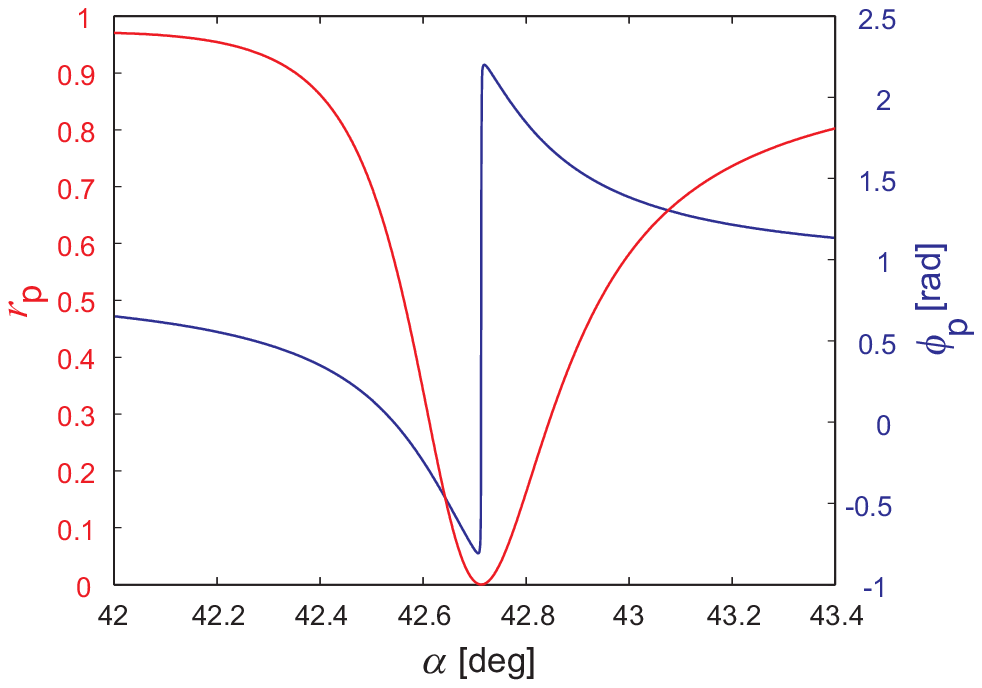}}}
\caption{Theoretical reflection coefficient $r_{p}$ and phase shift $\protect%
\phi _{p}$ of the reflected p-polarized light field. The incidence angle $%
\protect\alpha $ is varied around the surface plasmon angle at $42.7^{\circ
} $.}
\label{fig:refl_phase}
\end{figure}
The gold film on the prism surface has been fabricated by e-beam evaporation
and has a thickness of approximately $50\,\mathrm{nm}$. Below the gold film
a very thin layer of titanium has been added for improving adhesion of the
gold to the glass substrate. The prism is placed inside a vacuum chamber at
a base pressure of $5\cdot 10^{-9}\,\mathrm{mbar}$ and is mounted to a
temperature controlled heating. The partial pressure of Rb in the chamber is
controlled by a resistively heated dispenser and can be raised within a few
seconds up to several $10^{-6}\,\mathrm{mbar}$. Laser induced desorption of
Rb atoms can be investigated with an additional blue laser beam at a
wavelength near 406 nm and a maximum power of 20 mW (laser 2)\cite{klempt_ultraviolet_2006}.\newline

\section{Theoretical model}

The interpretation of our experiment is based on the observation that,
according to Brunauer, Emmet and Teller (BET), only fractions of a monolayer
can be adsorbed if the vapor pressure of an adsorbant is smaller than the
saturation vapor pressure $p<p_{\mathrm{sat}}$ \cite{brunauer38}. Thus the
surface coverage $\Theta $ only takes values between $0$ and $1$, i.e. $%
0<\Theta <1$. We model the temporal change of the surface coverage by the
rate equation 
\begin{equation}
\dot{\Theta}=-C\Theta +I_{\mathrm{in}}\sigma S(\Theta ),
\label{eq:rate_eq_1}
\end{equation}%
with a desorption constant $C$, an inpinging flux of atoms on the surface $%
I_{\mathrm{in}}$, a typical area per adsorbed atom of $\sigma =10^{-15}~%
\textrm{cm}^{2}$ and a sticking probability $S(\Theta )$. The atomic flux
hitting the surface is given by $I_{\mathrm{in}}=nv/4$ with the factor $4$
following from integrating the Knudsen cosine law over a half sphere, the
ideal gas vapor density $n=p/k_{B}T$ and the thermal velocity of the atoms $%
v=\left( 2k_{B}T/m\right) ^{1/2}$. For simplicity we define the flux per
pressure $R$ in units of adsorbed monolayers via $I_{\mathrm{in}}\sigma
\equiv Rp$, i.e. $R=\sigma \left( 1/8mk_{B}T\right) ^{1/2}$. The sticking
probability is taken from the Langmuir model $S(\Theta )=S_{0}\cdot
(1-\Theta )$, with the initial sticking probability $S_{0}$ \cite{Masel1996}%
. Note that $S(\Theta )$ vanishes for full coverage ($\Theta =1$). This
guaranties that only fractions of a monolayer can be adsorbed. With these
definitions (\ref{eq:rate_eq_1}) reads 
\begin{equation}
\dot{\Theta}=-(C+S_{0}Rp)\Theta +S_{0}Rp,  \label{eq:rate_eq_2}
\end{equation}%
In equilibrium ($\dot{\Theta}=0$) the surface coverage is given by 
\begin{equation}
\Theta _{\mathrm{eq}}=\frac{1}{1+C/S_{0}Rp}.  \label{eq:coverage_gg}
\end{equation}%
The equilibrium coverage depends on the pressure in the chamber. If the
pressure is rapidly changed the new equilibrium is exponentially reached
within an $1/e$-timescale 
\begin{equation}
\tau =\frac{1}{C+S_{0}Rp},  \label{eq:tau}
\end{equation}%
as can be deduced from (\ref{eq:rate_eq_2}).

\section{Observations}

\begin{figure}[tbph]
\centerline{\scalebox{0.9}{\includegraphics{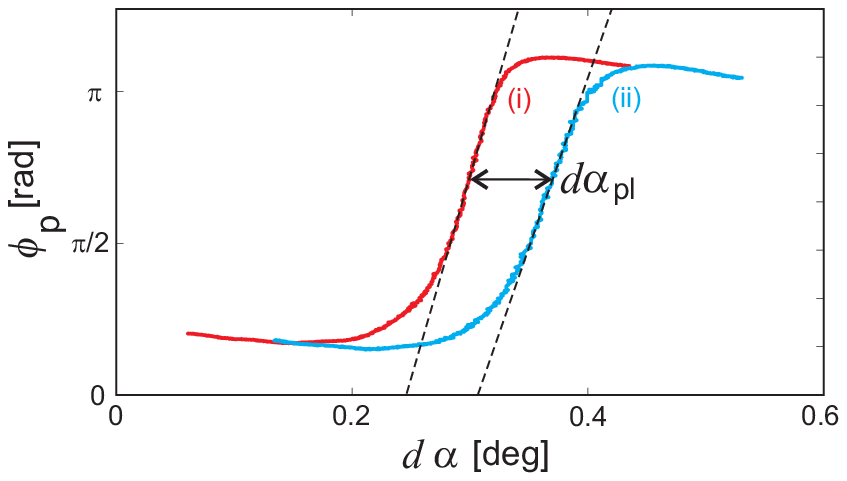}}}
\caption{Phase $\phi _{p}$ vs. incidence angle shift $d\protect\alpha$ before (i) and
after (ii) adsorption of Rb atoms.}
\label{fig:scans}
\end{figure}
We measure the adsorption of rubidium with the above mentioned setup by two
different methods. In the first method the incidence angle is scanned and
from the obtained signal the position of the plasmon angle is determined.
This method is applicable for a plasmon angle which changes only little
during the scan time of approximately one minute. Figure~\ref{fig:scans}
shows an example. The relation between the shift of the plasmon angle $%
d\alpha _{\mathrm{pl}}$ and the thickness of the adsorbed layer $dl$ is taken
from a four-layer matrix model (layer 1: glass substrate, layer 2: gold
film, layer 3: Rb adsorbate, layer 4: vacuum) which takes multiple
reflections at the boundaries into account \cite{deutsch95}. With the
dielectric constant of gold $\epsilon _{\mathrm{Au}}=-22.9+1.42\times i$ and
Rb $\epsilon _{\mathrm{Rb}}=-4+0.42\times i$ \cite{etchegoin06, smith70} the
result is $d\alpha _{\mathrm{pl}}/dl=0.1^{\circ }/\textrm{nm}$. The observed
shift of the plasmon angle in figure~\ref{fig:scans} of $d\alpha _{\mathrm{pl%
}}=0.072^{\circ }$ thus corresponds to an adsorbed Rb layer thickness of $%
l=0.72$~nm.\newline

\begin{figure}[tbph]
\centerline{\scalebox{0.8}{\includegraphics{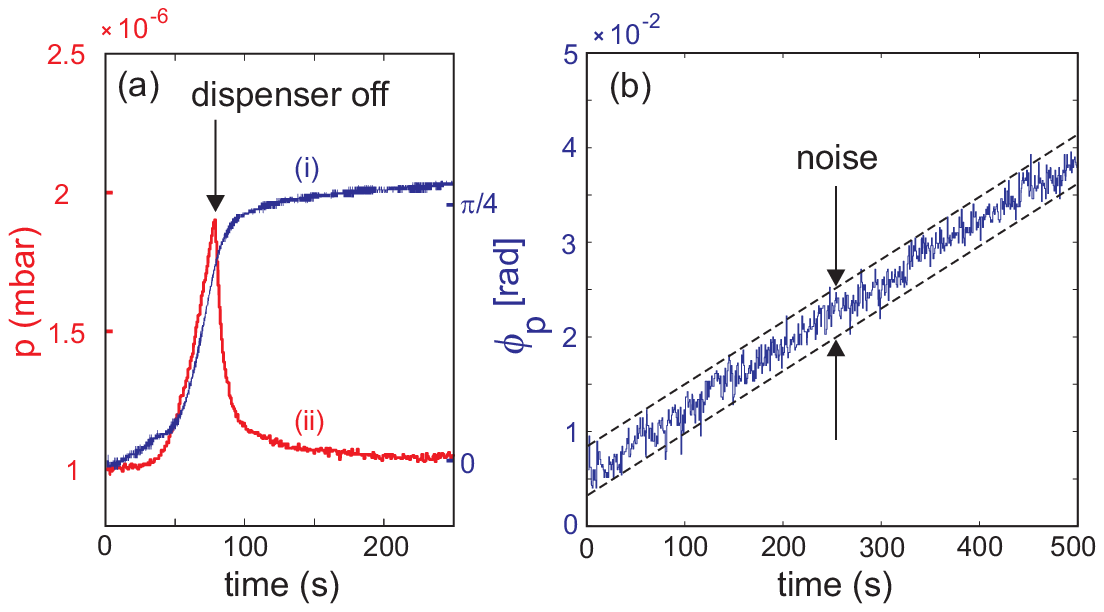}}}
\caption{(a) Observed phase shift (i) and Rb vapor pressure (ii) after
activating the Rb dispenser at $\textrm{time}=0$ for a period of 80s. (b) Estimation
of the experimental phase noise level.}
\label{fig:adsorption_phase}
\end{figure}

The second, more sensitive method records the phase $\phi _{p}$ with the
incidence angle held fixed at the plasmon angle. Adsorption and desorption
of atoms can now be observed in real time by looking at the temporal change
of the phase $\phi _{p}$. For not too large variations of $\alpha _{\mathrm{%
pl}}$ the relation between plasmon angle and phase is linear with a slope of 
$d\phi _{p}/d\alpha _{\mathrm{pl}}=42.1~$rad$/^{\circ }$ that is taken from
the left curve in figure~\ref{fig:scans}. Note that after adsorption of Rb (right curve) the
slope is with $d\phi _{p}/d\alpha _{\mathrm{pl}}=35.1~\textrm{rad}/^{\circ }$
slightly smaller. For large changes of the plasmon angle the working point
is shifted from the linear regime to a regime where the phase change
saturates. Thus the second method can only be applied for small changes of
the plasmon angle, unless the incidence angle is recalibrated. Typical data
are shown in figure~\ref{fig:adsorption_phase}. In (a) the Rb vapor pressure
is quickly increased by activating the dispenser for about 80s (curve (ii)).
The pressure is determined with a standard ion-gauge. Curve (i) shows the
resulting change of the phase $\Phi _{p}$. During the pulsed increase of the
vapor pressure the Rubidium coverage also grows and remains at an increased
value after the puls.  Figure~\ref{fig:adsorption_phase} (b)  shows a
similar measurement with higher resolution which allows to estimate the
sensitivity of the method. After low pass filtering the signal with an
integration time of $T=50$~ms the noise of the phase signal is on the order
of $\Delta \phi _{p}=5\times 10^{-3}$~rad.  At this noise level the
uncertainty of the measured layer thickness is $\Delta l=\Delta \phi
_{p}\cdot d\alpha _{\mathrm{pl}}/d\phi _{p}\cdot dl/d\alpha _{\mathrm{pl}%
}=1.2$~pm. With an average distance between two atoms given by twice the
covalent radius of Rb of $2r_{c}=440$~pm, the sensitivity amounts to better
than $\Delta l/2r_{c}=0.3\%$ of an adsorbed monolayer. This sensitivity
corresponds to a resolveable relative refractive index change of $\Delta n/n=2\cdot 10^{-6}$. 

\begin{figure}[tbp]
\centerline{\scalebox{0.9}{\includegraphics{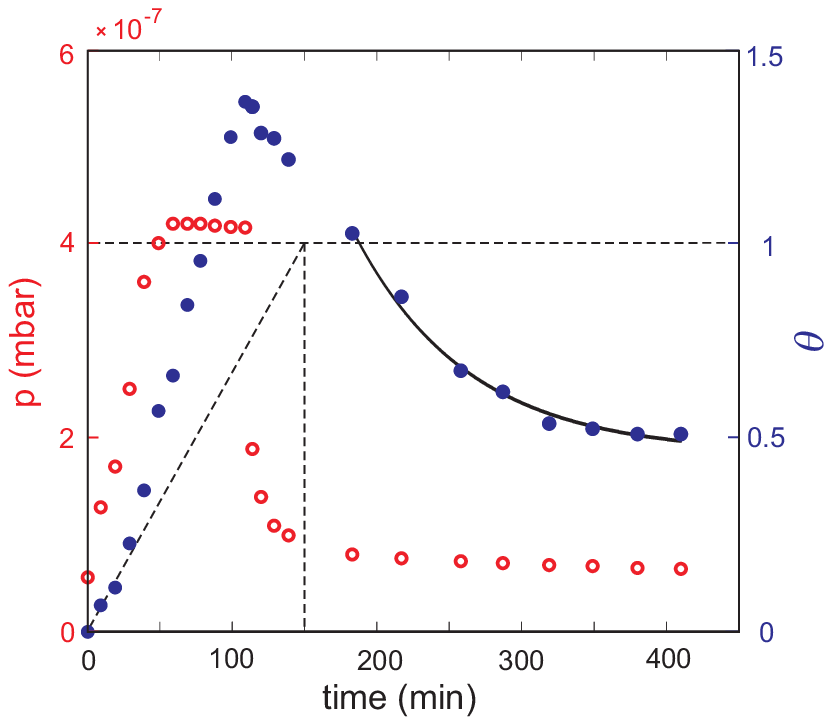}}}
\caption{Observed monolayer coverage (blue solid dots, right vertical axis)
during and after a temporal increase of the Rb vapour pressure  (red open
circles, left vertical axis). From the linear coverage growth during the
first 20 minutes (black doted line) the growth time per monolayer can be
deduced. The decay of the coverage after the pulse is fitted with an
expontial function.}
\label{fig:desorption_angle}
\end{figure}

In the experiment the gold surface is heated to a temperature of $T_{S}=68^{\circ }$~C . The resulting Rb saturation vapor pressure of $p_{%
\mathrm{sat}}=2\cdot 10^{-5}$~mbar is one order of magnitude larger than the
maximum Rb vapor pressure in the experiment of $p_{\mathrm{max}}=2\cdot
10^{-6}$~mbar. The above assumption of a fractional coverage is thus well
justified. In figure~\ref{fig:desorption_angle} the observed coverage $%
\Theta $ is plotted during and after a temporal increase of the Rb vapor
pressure. During the first 20 min the pressure stays very much near a value of about $p=1\cdot 10^{-7}$~mbar.  The coverage
growth rate at this pressure can be derived by linearly extrapolating the
coverage increase with time. We obtain  $d\Theta /dt=\left( 150\,\textrm{min}%
\right) ^{-1}$. Putting this value into (\ref{eq:rate_eq_2}) and setting $\Theta=0$ the initial sticking coefficient can be determined
to be $S_{0}=0.8\%$. After $200$ min we reduce the pressure back to a value
of $p=7\cdot 10^{-8}$~mbar. Now the atoms thermally desorb and the coverage
decays exponentially. The thermal desorption constant $C=C_{\mathrm{th}}$
can be determined from (\ref{eq:tau}) by measuring the 1/e decay time $\tau $%
. Fitting an exponential curve to the data yields $\tau =4.7\cdot 10^{3}$~s.
For the desorption constant we obtain $C=1/\tau -S_{0}Rp=1.3\cdot 10^{-4}~%
\textrm{s}^{-1}$. In order to show the consistency of the model with the
determined parameters the equilibrium coverage is calculated from (\ref%
{eq:coverage_gg}) to be $\Theta _{\mathrm{eq}}=0.4$. This value is
comparable to the observed value in figure~\ref{fig:desorption_angle} of $%
\Theta _{\mathrm{eq}}=0.5$.

\begin{figure}[tbph]
\centerline{\scalebox{0.9}{\includegraphics{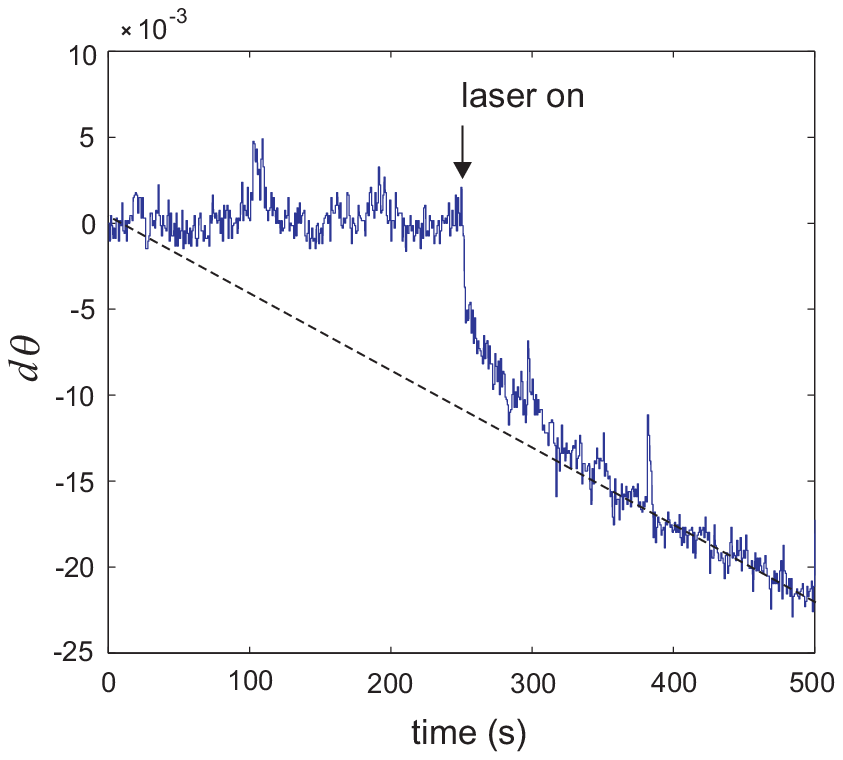}}}
\caption{Laser-induced desorption of Rb. If the surface is exposed to blue
laser light, the surface coverage starts to decrease.}
\label{fig:desorption_laser}
\end{figure}

The equilibrium coverage can be further reduced by exposing the surface to a laser
beam with a wavelength near $406\,\mathrm{nm}$ ($20\,\mathrm{mW}$
laserpower, beam diameter $1$~mm). Figure~\ref{fig:desorption_laser} shows
the resulting change of the surface coverage $d\Theta $. As soon as the blue
laser is turned on, the coverage decreases in time with a rate of $\dot{%
\Theta}=-4.4\cdot 10^{-5}~\textrm{s}^{-1}$. The total desorption coefficient
is now the sum of the thermal and the laser-induced desorption coefficient $%
C=C_{\mathrm{th}}+C_{\mathrm{li}}$. With (\ref{eq:rate_eq_2}) and for $\Theta =0.5$
and a pressure of $p=2\cdot 10^{-7}$~mbar the total desorption coefficient
amounts to $C=3.1\cdot 10^{-4}~\textrm{s}^{-1}$. This results in a pure
laser-induced desorption rate of $C_{\mathrm{li}}=1.8\cdot 10^{-4}~\textrm{s}^{-1}$ at a laser power of $20\,\mathrm{mW}$. The new equilibrium coverage
of $\Theta _{\mathrm{eq}}=0.42$ is reached witin a $1/e$~time of $\tau
=1.9\cdot 10^{3}$~s.

%%conclusion

\section{Conclusion}

We have applied a recently introduced method \cite{knig_real-time_2008} for
detecting de- and adsorption phenomena by means of phase-sensitive detection
of a surface plasmon resonance to Rubidium on a gold surface. Adsorption and
desorption of Rb were observed and interpreted with a rate model. We reach a
sensitivity of better than $0.3\%$ of a monolayer, respectively a refractive
index unit change of $\Delta n=2\cdot 10^{-6}$~RIU within an integration
time of $50$~ms. This is comparable to typical sensitivities reached with
surface plasmon resonance sensors \cite{gauglitz99}. The observed
sensitivity in this paper is limited by technical noise which leaves much
room for substantial improvement, e.g. by a shot-noise limited detection
scheme, by suppression of laser frequency noise, and by longer integration
time. As has been shown in \cite{knig_real-time_2008}, a resolution of $%
\Delta n=2\cdot 10^{-8}$~RIU within an integration time of $0.1$~s seam
feasible even far above the shot-noise limit. Such a sensitivity exceeds the
best values reported for converntional detectors, e.g. in \cite{grigorenko09}%
. By comparing the experimental data with a Langmuir type adsorption model
we find a desorption constant of $C=1.3\cdot 10^{-4}~\textrm{s}^{-1}$ and an
initial sticking probability $S_{0}=0.8\%$. Laser-assisted desorption has
been observed. A desorption coefficient of $C_{\mathrm{li}}=1.8\cdot 10^{-4}~%
\textrm{s}^{-1}$ is determined for a laser power of $20$~mW.\newline

%Acknowledgement
\ack We acknowledge financial support by the DFG within the EuroQuasar
program of the ESF and thank A. Hemmerich for inspiring discussions.

\end{document}